\documentclass[acmlarge]{acmart}

\usepackage{placeins}

\usepackage{iftex}
\ifPDFTeX
  \usepackage[utf8]{inputenc}
  \usepackage[T1]{fontenc}
  \usepackage{CJKutf8}
  \AtBeginDocument{\begin{CJK*}{UTF8}{gbsn}}
  \AtEndDocument{\end{CJK*}}
\else
  \usepackage{fontspec}
  \usepackage{xeCJK}
  \setCJKsansfont{FandolHei-Regular.otf}[
    BoldFont = FandolHei-Bold.otf,
  ]
  \setCJKmonofont{FandolFang-Regular.otf}
  \usepackage{unicode-math}
\fi

\usepackage[labelsep=endash]{caption}

\usepackage[many]{tcolorbox}
\usepackage{varwidth}
\usepackage{environ}
\usepackage{xparse}
\microtypesetup{protrusion=true} % tweak as desired
\UseMicrotypeSet[protrusion]{basicmath}          % note: spelled "protrusion"
\usepackage[format=plain,justification=centering]{caption}

\usepackage{newunicodechar}

\usepackage{tikz}
\usetikzlibrary{arrows.meta,positioning}
\usetikzlibrary{calc}

\usepackage{libertinus}
\newunicodechar{’}{\textquotesingle}
\newunicodechar{′}{{\ttmath ′}}
\usepackage{changepage}

\usepackage{slantsc}

\setcopyright{rightsretained}
\copyrightyear{2025}
\acmYear{2025}
\acmDOI{10.XXXX/XXXXXXX.XXXXXXX}

\copyrightyear{2026}
\acmYear{2026}
\setcopyright{rightsretained}
\acmConference[PLoP 2025]{32nd Conference on Pattern Languages of Programs, People, and Practices}{October 12--15, 2025}{Skamania Lodge, Columbia River Gorge, Washington, USA.}
\acmISBN{978-1-941652-22-0}
\acmDOI{10.64346/PLoP2025p24}

\usepackage{graphicx}
\usepackage{grffile}
\usepackage{longtable}
\usepackage{array}
\usepackage{calc}

\usepackage{wrapfig}
\usepackage{rotating}
\usepackage[normalem]{ulem}
\usepackage{textcomp}
\usepackage{capt-of}
\usepackage{hyperref}
\usepackage{mdframed}
\usepackage{afterpage}

\usepackage{booktabs}
\usepackage{tabto}

\usepackage[inline]{enumitem}

\usepackage[pagewise]{lineno}

\usepackage{natbib}
\usepackage{float}
\usepackage{tikz}

\usepackage{placeins}
\usepackage{bold-extra}

\usepackage{enotez}
\renewcommand{\endnote}[1]{}

\newcommand{\markbf}[1]{\textsuperscript{\textbf{#1}}}
\setenotez{counter-format = alph, mark-cs = \markbf}
\DeclareInstance{enotez-list}{sverre}{paragraph}{heading={},notes-sep=\baselineskip,format=\normalsize\normalfont\raggedright\leftskip1.8em,number=\makebox[0pt][r]{#1.\ }\ignorespaces,}
\usepackage{epigraph}
\date{}
\title{Patterns of Patterns III: Agentic Patterns}
\hypersetup{
 pdfauthor={Joseph Corneli et al.},
 pdftitle={Patterns of Patterns III},
 pdfkeywords={},
 pdfsubject={},
 pdfcreator={Emacs 30.0.50 (Org mode 9.6.1)},
 pdflang={English}}

\title{Patterns for a New Generation: AI and Agents}

\author{Joseph Corneli}
\authornote{Corresponding author, jcorneli@brookes.ac.uk.}
\email{jcorneli@brookes.ac.uk}
\orcid{1234-5678-9012}
\affiliation{%
  \institution{Oxford Brookes University}
  \streetaddress{Gipsy Lane}
  \city{Oxford}
  \country{UK}
  \postcode{OX3 0BP}
}
\affiliation{%
  \institution{Hyperreal Enterprises Ltd}
  \streetaddress{272 Bath Street}
  \city{Glasgow}
  \country{UK}
  \postcode{G2 4JR}}

\author{Charles J. Danoff}
\affiliation{%
  \institution{Mr Danoff’s Teaching Laboratory}
 \streetaddress{PO Box 802738}
 \city{Chicago}
 \state{IL}
  \country{USA}
  \postcode{60680}}
\email{contact@mr.danoff.org}

\author{Raymond S. Puzio}
\email{rsp@hyperreal.enterprises}
\affiliation{%
  \institution{Hyperreal Enterprises Ltd}
  \streetaddress{272 Bath Street}
  \city{Glasgow}
  \country{UK}
  \postcode{G2 4JR}}

\author{Sridevi Ayloo}
\email{sayloo@citytech.cuny.edu}
\affiliation{%
  \institution{City Tech}
  \streetaddress{300 Jay St}
  \city{Brooklyn}
  \state{NY}
  \country{USA}
  \postcode{11201}
}

\author{Sergio Belich}
\email{serge.belich85@citytech.cuny.edu}
\affiliation{%
  \institution{City Tech}
  \streetaddress{300 Jay St}
  \city{Brooklyn}
  \state{NY}
  \country{USA}
  \postcode{11201}
}

\author{Andre Wilkinson}
\email{Andre\_w3@yahoo.com}

\affiliation{%
  \institution{City Tech}
  \streetaddress{300 Jay St}
  \city{Brooklyn}
  \state{NY}
  \country{USA}
  \postcode{11201}
}

\author{Mary Tedeschi}
\email{mtedeschi@pace.edu}
\affiliation{%
  \institution{Pace University}
  \streetaddress{One Pace Plaza}
  \city{New York}
  \state{NY}
  \country{USA}
  \postcode{10038}
}
\author{Pauline Mosley}
\email{pmosley@pace.edu}
\affiliation{%
  \institution{Pace University}
 \streetaddress{One Pace Plaza}
 \city{New York}
 \state{NY}
  \country{USA}
  \postcode{10038}}

\renewcommand{\shortauthors}{Corneli et al.}

\begin{abstract}
Design patterns have been used in various fields of inquiry and endeavour to externalize procedural knowledge in a form that supports human reasoning and coordination. In this paper, we show that contemporary Large Language Model (LLM)-based systems can also read, generate, and reason with design patterns written in a structured template. We describe an experimental workflow in which patterns function as shared priors for action selection, reflection, and revision in hybrid human/agent settings.  Drawing on the Active Inference Framework, we illustrate how patterns can guide agent behavior without fully prescribing it. This provides a proof of concept that pattern-capable agents can be created using now-standard software tools. We  discuss implications for software development, education, business, and AI governance.

\end{abstract}

\ccsdesc[500]{Human-centered computing~Collaborative and social computing}
\ccsdesc[500]{Social and professional topics~Computing education}
\ccsdesc[300]{Applied computing~Education}
\ccsdesc[300]{Software and its engineering~Design patterns}
\ccsdesc[300]{Computing methodologies~Artificial intelligence}

\makeatletter
\def\@ACM@checkaffil{% Only warnings
    \if@ACM@instpresent\else
    \ClassWarningNoLine{\@classname}{No institution present for an affiliation}%
    \fi
    \if@ACM@citypresent\else
    \ClassWarningNoLine{\@classname}{No city present for an affiliation}%
    \fi
    \if@ACM@countrypresent\else
        \ClassWarningNoLine{\@classname}{No country present for an affiliation}%
    \fi
}
\makeatother

\begin{document}
\input{HillsidePatch-doi.tex}
\maketitle

\addtocounter{section}{-1}
\section{Definition}\label{sec:def}
A \emph{design pattern} is a learned generalization arising from a review process that describes how intentions, actions, and outcomes tend to relate across situations.  A candidate pattern becomes a pattern only when it has survived repeated enactment-and-review cycles in which its failure modes are recorded and used to revise it.

This is not how people talked about design patterns when they were introduced \cite{alexander1970a,alexander1977a} or popularised \cite{gamma1994a,alexander1999a}, but the description naturally applies to that foundational work.  The difference here is that we focus less on patterns as methods \cite{leitner2015pattern}, and more on how they are learned in the first place. This framing introduces a potential regress. If patterns are learned through action review, and the reviews themselves require competent performance, how is pattern-theoretic competence developed?  Not through the application of further patterns, we assert, but as a practice of care that enables judgment without reducing it to rule-following.  If the regress was followed further, it would bottom out in adaptive pattern-sensitive mechanisms of the sort described by Hawkins \citeyearpar{hawkins2021thousand}—or, in computational processes that bridge the symbolic and subsymbolic realms, like the ones described in this paper.

\section{Introduction}
Decisions about technology are shaped largely by how humans communicate, explain, justify, and contest what computers do. The associated intentions are often embedded in code, models, or opaque interfaces, where they turn into operational definitions and decision criteria. These are typically beyond the reach of non-programmers, who are trained to reason through text, argument, and interpretation. This paper examines whether design patterns—as defined above, and as recorded, typically, in the form of structured, narrative descriptions of recurring problems and responses—can serve as a shared language through which both humans and AI agents reason about action.  This would enable non-programmers to participate more fully in the design, critique, and governance of computational systems.

Although there have been previous initiatives that engaged with design patterns
using computational tools (such as the first-ever wiki, \texttt{c2.com}), the kind of project we are interested
in really only becomes feasible to explore with the introduction of
Large Language Models (LLMs).  The work we present is related to
several existing technical practices that developed in that setting:
``context engineering'' \cite{contextwindow, contextbuddy}, ``chain-of-thought scaffolding'', and ``spec-driven development''.\footnote{\url{https://deliberate.codes/blog/2026/spec-driven-development-an-introduction/}}  In other
words, respectively,
\begin{itemize}
\item structuring prompts, system instructions, and retrieved context to shape LLM behavior;
\item steering how computational reasoning unfolds over time; and,
\item understanding specifications—rather than code—to be the primary programming artifact.
\end{itemize}
Design patterns have some particularly nice properties in this
setting: they persist over time and across sessions, and they support
progressive work and verification at various levels of formality.

% connect context window to "1. Changing context as a decentered center." from https://ceur-ws.org/Vol-739/paper_5.pdf
% refernece https://mrhillsman.com/posts/context-engineering-realized-context-window-architecture/ and connect it to layers of CLA

It is convenient to begin with the anatomy of the patterns we use,
which is a light adaptation of the standard design pattern template.  The
template is important because it is an interface between human
reasoning, LLM generation, and logic-based verification.  The fields we use are as follows: 

\textbullet\ Patterns begin with a \textbf{summary}, which can
express a short description of a \emph{feature} to be
built.  We can use patterns to build software or
other resources, including for example, an argument
or a way of working.  The feature might be a
\emph{conclusion} that the reader should draw from
reading the pattern.  % The summary summarizes the further fields: \textbf{context}, \textbf{if}, \textbf{however}, \textbf{then}, \textbf{because}, \textbf{next-steps}.

\textbullet\ The \textbf{context} characterizes a sphere of endeavour.

\textbullet\ The pair \textbf{if}$\leftrightarrow$\textbf{however}
outlines a tension that exists within the context.

\textbullet\ The \textbf{then} clause suggests
actions to take to improve the situation, though we
don't require that the actions fully resolve the
tension: they ameliorate or help manage it;
the \textbf{because} clause gives the rationale; an
optional subfield \textbf{evidence} can collect
specific pieces of evidence.

\textbullet\ The \textbf{next-steps} outline one or
more steps to take to give evidence that would
strengthen the validity of the pattern itself.

Some examples of design patterns written in this format are given in
Box \ref{box:3} in the Appendix.  Relative to most prior work on design patterns, these have the
novel quality of being \emph{machine generated}, using a Large
Language Model based agent.  The paper will explore some of the
implications of working with design patterns using LLMs—both for
guiding the behavior of the agents, and organizing hybrid
(human+agent) interactions.  This way of working is different from
using design patterns to control generation in a unitary way—given
design pattern {\sffamily A}, build software {\sffamily B} that has
all the nice properties that {\sffamily A} specified.  At least
potentially, design patterns could govern the entire process of
software generation.  However, that is a complex problem, and we do
not claim to have solved it. Rather, in this short paper, we
illustrate some of the thinking and examples that could inform further
exploration of the use of software patterns together with AI agents. 

% add a bit of forward refs about teaching

The remainder of the paper is organized as follows. Section~\ref{sec:peeragogy}
briefly introduces peeragogy and the pattern traditions that inform this work.
Section~\ref{sec:system-overview} outlines the experimental prototype we built.
Section~\ref{sec:pattern-selection} describes the way that system selects and uses patterns.
Section~\ref{sec:aif} describes the cognitive science concepts that underpin that process.
Section~\ref{sec:related-work} describes related work, and
Section~\ref{sec:study} and Section~\ref{sec:pvcs} describe two
practical provocations related to applications in teaching and enterprise.
Section~\ref{sec:conclusion} presents our conclusions.

\section{Background}\label{sec:peeragogy}
Peeragogy is a peer-produced and peer-facilitated approach to learning
in which participants collaboratively develop goals, roles, and
artifacts, and iteratively refine their practices through reflection
and coordination.  In this sense, peeragogy is both a theory of
learning and an operational method for organizing shared inquiry and
production \cite{corneli2015patterns,peeragogy-handbook}.  It sits
alongside longstanding research on topics such as communities of
practice and peer production
\cite{wenger1998communities,benkler2006wealth}, articulating the
learning-specific aspects thereof.  As such, peeragogy offers a
natural way to think about much existing design pattern practice
(e.g., writers' workshops are a peer produced peer learning space
\cite{gabriel1writer}).  In this paper, we draw our experience working
together peeragogically to motivate how patterns can serve as shared
artifacts that help humans and agents align on intent,
evidence, and next steps.

\section{System Overview}\label{sec:system-overview}

In practice, it is relatively simple to get contemporary chatbots to produce design
patterns in the format we listed in the Introduction.  Indeed, due to the way such
agents work, if you want to try this, the steps that we recommend are
(1) provide the pattern template as a system prompt or custom
instruction; (2) tell the agent \emph{not} to use patterns by default,
or it probably will use them everywhere; (3) request patterns
explicitly when you want them.  (See Appendix, Box \ref{box:1}
for a viable prompt.)  The fact that it is now easy to ``do it
yourself'' and play around with design patterns and LLMs resonates
with Hofstadter's invitation in \emph{G\"odel, Escher, Bach} to engage
actively with formal systems \cite{hofstadter1999godel}.  That being
said, simply creating patterns that follow a template does not
guarantee that those patterns will be any good.  Based on our
reflections in previous papers
\cite{corneli2021patterns,corneli2022patterns}, we think that there
are three primary uses for patterns.  Patterns can help:

\begin{enumerate}[label=\Alph*.]
\item \textbf{coordinate perspectives} through a process of observation and argumentation,

\item \textbf{support decision making} by relating conditions to possible actions and alternatives,

\item \textbf{collect evidence for their own salience} through a process of testing and revision.
\end{enumerate}

These, effectively, become design requirements.  How we go about doing
these things depends on the application area and the technology used.
Patterns that are mediated by paper (Alexander et al.)~have a
different feel from patterns that are mediated by wikis (Cunningham et
al.)\footnote{\url{https://c2.com/}}.  When thinking about how
patterns can be used together with AI agents, one key choice that
needs to be made is the degree to which design patterns will
\emph{regulate} or \emph{guide} agent behaviour.  Consider, for
example, that issue trackers and version control systems guide, but do
not regulate human computer programmers, whereas things like contract
and copyright law do, ultimately, have regulating effects.
We attempted to use patterns as a mechanism
for guiding rather than controlling coding agents (such as
OpenAI's Codex\footnote{\url{https://chatgpt.com/features/codex/}}, Anthropic's Claude Code,\footnote{\url{https://code.claude.com/docs/en/overview}} etc.).  Figure~\ref{fig:loop} shows the conceptual framework.

\begin{figure}[h]
\begin{center}
% TikZ diagram: AIF engine components
% Requires: \usepackage{tikz}
% Optional: \usetikzlibrary{arrows.meta,positioning}
\begin{tikzpicture}[>=Latex, font=\small]
  \tikzset{
    frame/.style={draw, rounded corners, minimum width=100mm, minimum height=40mm},
    box/.style={draw, rounded corners, align=center, minimum width=26mm, minimum height=10mm},
    arrow/.style={-Latex}
  }

  \node[frame] (outer) {};
  \node[anchor=north, font=\small\bfseries] at (outer.north) {AIF Engine};

  \node[box] (gen) at ($(outer.center)+(-32mm,5mm)$) {Generative\\Model\\\footnotesize (Patterns)};
  \node[box] (act) at ($(outer.center)+(0mm,5mm)$) {Action\\Selection\\\footnotesize (PSR)};
  \node[box] (upd) at ($(outer.center)+(32mm,5mm)$) {Belief\\Update\\\footnotesize (PUR)};
  \draw[arrow] (gen.east) -- (act.west);
  \draw[arrow] (act.east) -- (upd.west);

  \node[below=10mm of gen, font=\footnotesize] (cap-gen) {Pattern lib as priors};
  \node[below=10mm of act, font=\footnotesize] (cap-act) {Which pattern now?};
  \node[below=10mm of upd, font=\footnotesize] (cap-upd) {Did it work?};

  \draw[arrow] (gen) -- (cap-gen);
  \draw[arrow] (act) -- (cap-act);
  \draw[arrow] (upd) -- (cap-upd);
\end{tikzpicture}
\end{center}
\caption{Pattern-driven scheme for guiding coding agents\label{fig:loop}}
\end{figure}

What we have implemented is a workflow for observing and guiding pattern-competent agents inside an experimental workspace we call FuLab.  The ``fu'' morpheme stems from our engagement with future-directed thinking, but may also evoke \emph{kung fu} (功夫)---skill achieved
through practice and discipline---as well as \emph{butoh-fu}
(舞踏譜)---a genre of evocative prompts for the butoh dance
form.\footnote{\url{https://butoh-kaden.com}} We have created
\texttt{fuclaude} and \texttt{fucodex} CLI wrappers for the ``off the
shelf'' Claude and Codex agents, as well as a \texttt{fubar.el} Emacs
package, which humans can use both to observe and experience the agents' workflow firsthand. The code is available at \url{https://github.com/tothedarktowercame/futon3}.

\section{Pattern Selection and Use}\label{sec:pattern-selection}

The concept of effective behavior involves reasoning from intention to action. This can be done in a forward direction with planning, or retrospectively, through review.  The approach we consider in the current paper invites coding agents to select design patterns that operationalize the user's stated intention as they run.

%% \begin{itemize}
%%     \item Understanding effective behavior involves reasoning from the intention to action.
%%     \item This can happen retrospectively, but that's not so easy to do in real time as a system is running!
%%      \item Instead, selecting patterns that can help operationalize the intention in a forward direction.
%%      \item Coding agents can easily go off the rails if this overall way of working isn't secured.
%% \end{itemize}

In order to encourage the coding agent to work in a way that matches
this workflow, we give it the following help text when
it starts a new coding session (the tools define Remote Procedure Calls (RPC) and are available as command-line ``helpers''):
\begin{quote}
\begin{verbatim}
Tool roster (you should emit these signals when you do the corresponding action)
- pattern-select library/<pattern>    <state why you want to read it>
- pattern-use    library/<pattern>    <state where you will apply it>
- musn-plan      <outline your plan>
- wide-search    ; an alias for rg
\end{verbatim}
\end{quote}

The workflow is driven through a CLI runner that streams Codex JSON,
appends session events, and emits records at turn
boundaries. Resuming a session appends to the same trace. At present, anchor
resolution is limited to the session's event log; verification against
external artifacts (e.g., code diffs or tests) remains future work. 

\section{Patterns and Active Inference}\label{sec:aif}
Christopher Alexander introduced a star-rating that encodes confidence in each pattern that was supported by real-world evidence.  Here, we use a similar idea to derive a confidence rating from machine-checkable properties of the pattern text.  There are two questions: how broadly applicable is the pattern, and how reliable is it within its domain of application?  For now, we conflate these two into one \emph{precision prior} associated with the pattern's maturity state.

\begin{center}
\begin{tabular}{lllr}
\textbf{yes} next-steps + \textbf{yes} evidence & $\rightarrow$ & \texttt{:active} & precision prior: 0.8 \\
\textbf{yes} next-steps + no evidence & $\rightarrow$ &  \texttt{:greenfield} & precision prior: 0.4\\
no next-steps + \textbf{yes} evidence & $\rightarrow$ &  \texttt{:settled} & precision prior: 0.9\\
no next-steps + no evidence & $\rightarrow$ &  \texttt{:stub} & precision prior: 0.2
\end{tabular}
\end{center}

This makes mature patterns slightly more likely to be sampled, although \texttt{:stub} patterns can be selected when the system is in an ``exploratory mode'', allowing new patterns to accumulate evidence through use.

To model this, we draw on ideas from cognitive science which themselves adapt concepts from thermal physics \cite{friston2019freeenergyprincipleparticular}.  In somewhat general terms, the idea is that the steady state that a system will arrive in, assuming it isn't exchanging with its environment, is predicted by maximising entropy; however, if the system can exchange heat with its surroundings, then the equilibrium is governed by something called expected free energy—which basically means considering the entropy of the bigger system, which is understood through a generative model.  The particular cognitive science interpretation of these ideas that we are using is the Active Inference Framework (AIF) \cite{FRISTON2016862}; in our setting, the generative model is a policy layer that governs pattern selection and belief update.   

The AIF layer implements stochastic policy sampling: given candidate patterns and their expected free energy scores ($G$), it computes softmax probabilities over $-G/\tau$ and samples a suggestion. The precision parameter $\tau$ (derived from uncertainty in the current context) controls the exploration--exploitation trade-off: low $\tau$ yields greedy selection of the best-scoring pattern (this corresponds to highly ordered states); high $\tau$ spreads probability across candidate patterns (this corresponds to more fluid states). Figure~\ref{fig:psr-pur} shows a \texttt{fucodex} run that surfaces PSR and PUR discipline; Box \ref{box:2} in the Appendix gives brief further details on the agents' behaviour.

\begin{figure}
\includegraphics[width=\textwidth]{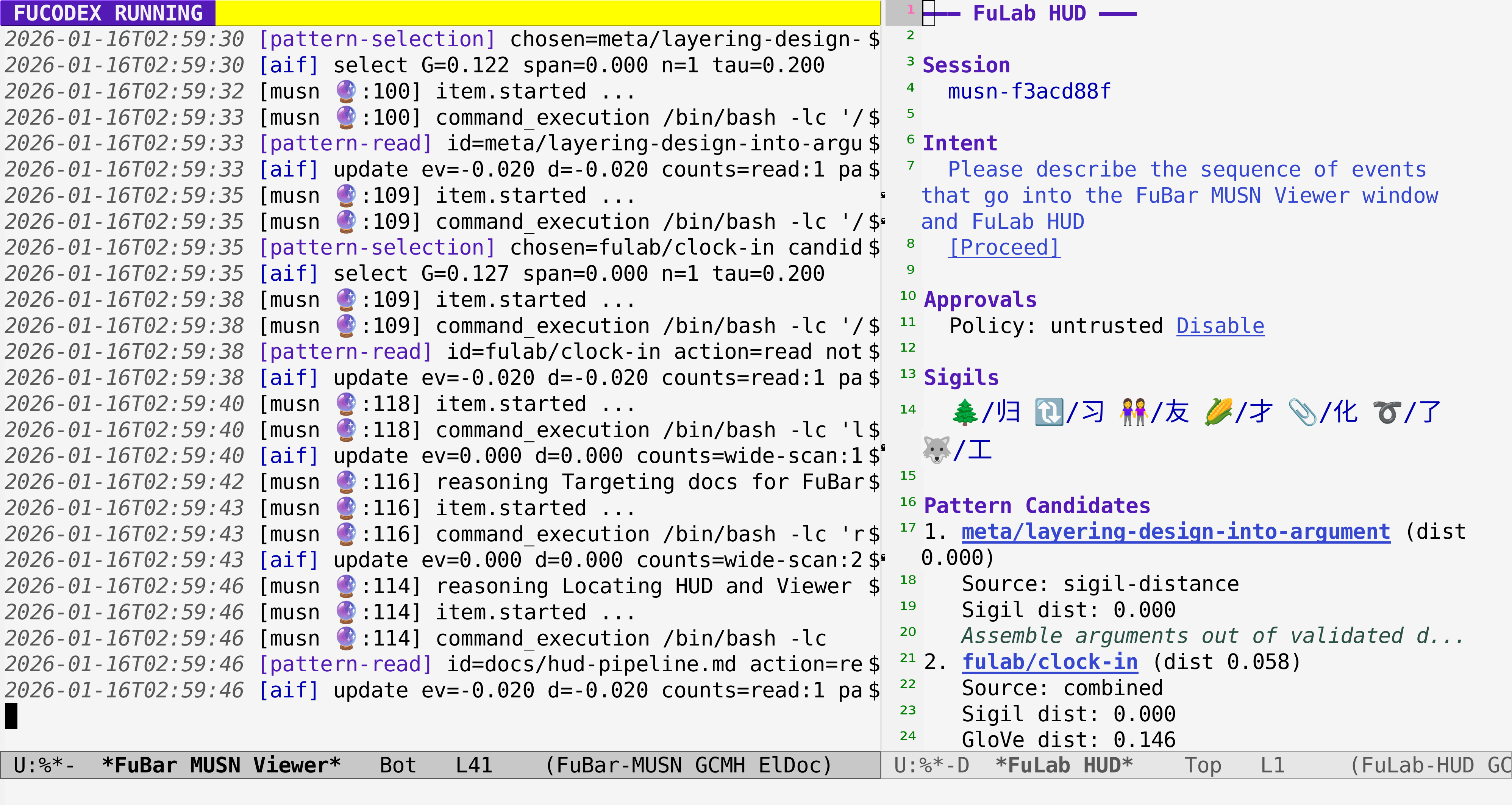}
\caption{\texttt{fucodex} run showing aspects of pattern selection and use}
\label{fig:psr-pur}
\end{figure}

%% \begin{itemize}
%% \item Limitations: simulation bias, lack of classroom pilots, narrative inference.
%% \item Commitments: richer "next steps" instrumentation and hybrid workshops.
%% \item Ethical framing: keep simulated results speculative and collaborative.
%% \item Pre-empt the skeptic: this work does not depend on AI agents, and patterns can also audit AI behavior.
%% \item Scope warning: the paper proposes a method with sketches, not hard evidence of efficacy.
%% \item Invite readers to explore the simulated-to-real transition responsibly.
%% \end{itemize}

%% \subsection{Simulated workshops}

%% \begin{itemize}
%% \item Use ChatGPT as a research catalyst to explore computationally tractable pattern use.
%% \item Treat simulations as design probes, not ground truth.
%% \end{itemize}

\section{Related Work}\label{sec:related-work}

While this paper does not rely directly on our earlier work with design patterns (summarized in Section~\ref{sec:peeragogy}), those efforts have had an indirect influence on the way the current project has evolved, and it is useful to give a brief account of how the earlier work relates to our current effort.

Patterns have been part of our work in the Peeragogy project for over a decade \cite{corneli2015patterns,peeragogy-handbook}. Inspired partly by that experience, we wrote two papers developing patterns for \emph{collaborative work with patterns} \cite{corneli2021patterns,corneli2022patterns}. We had the optimistic view that these methods—which draw on several existing mechanisms for scaffolding learning—could be quickly, easily, and reliably shared with others to build their capability for working with patterns.

Evidence from a series of workshops conducted at PLoP conferences suggests that the combined cognitive and practice demands of the frameworks we used are unlikely to be meaningfully assimilated within a single 60 to 90 minute conference session, and instead benefit from sustained enactment, guided reflection, and repeated use over time. We had more tangible success exploring the material in semester-long courses with postgrad students. By and large, we found that it was harder than we had anticipated to teach methods for working with patterns. The methods that we were working with were themselves complex multi-step frameworks developed by others that we had reworked for our context. Elaborating these methods into a set of design patterns \cite{corneli2022patterns} gave us more procedures to try in workshops, but this did not seem to lead to more robust insights or ways of working for participants. Going deep on one method bypassed the others; presenting methods in a light-weight way risked diluting learning. In semester-long courses, students had the opportunity to immerse themselves in the material, as evidenced by student projects that successfully applied the methods as well as thoughtful group reflections on the methods themselves. However, relative to our aim of finding a \emph{rapid} way to teach people effective pattern methods, we had reached a plateau.

Because the current paper draws some of its key inspirations from that earlier work, it is worth recapitulating the main ideas and landmarks.  With the PLACARD design pattern \cite{corneli2021patterns}, we aligned several existing social intelligence methods into a framework for working with design patterns. There were: the Project Action Review's five reflective review questions (PAR), Causal Layered Analysis's four layers for unpacking complex meanings and identifying possibilities for change (CLA), and Design Pattern Languages themselves, with their familiar three-part context-problem-solution schema (DPL). PLACARD organized these components into a methodology for understanding, reflecting on, and redesigning projects. In a follow-up paper \cite{corneli2022patterns}, we elaborated PLACARD into a design pattern language that added further methodological details, and we applied and evolved these methods in our workshops. Participants often found these workshops interesting and useful  \cite{10.1145/3698322.3698444}, however, even the ``core'' methods were relatively complex and not so easy to teach \cite{ayloo2024, Danoff2024}.
Our foregoing reflections on the three main uses for patterns  (Section \ref{sec:system-overview}) condense these methodological excursions in a jargon-free way. While we have not demonstrated that these condensed concepts are easy to teach \emph{per se}, we have seen they can be given a relatively straightforward computational interpretation.

In particular, the PSR (Pattern Selection Record) corresponds to the front half of a PAR (\emph{What is the current intention? Which pattern do we believe applies, and why now?}). The PUR (Pattern Use Record) corresponds to the back half (\emph{What happened when we applied it? What evidence did we observe? How should this affect future use?}). The main difference between PARs and PSRs/PURs is that PARs are typically applied retrospectively, whereas pattern-selection-and-use happens while a session runs.

A related attempt to coordinate activities across AI agents was seen in TheAgentCompany \cite{theagentcompany}. The authors worked with professionals to develop a list of tasks that humans working in different roles across departments in a software company need to do to complete their work. Next, they asked AI agents to do those tasks and saw an approximately 30\% completion rate.

Recent work on stochastic methods for establishing pattern sequences resonates with our use of AIF. Matovi\v{c} et al.~describe stochastic processes for constructing pattern sequences in organizational contexts \cite{matovic2025stochastic-processes-organizational}. Matovi\v{c} and Vrani\'{c} additionally explore stochastic trees and Bayesian belief networks for pattern sequencing in security patterns \cite{matovic2025stochastic-trees-bbn}, and present a related EuroPLoP study on stochastic processes for security patterns \cite{matovic2024stochastic-processes-security}. These approaches complement our focus on agent-mediated pattern selection by offering probabilistic formalisms for sequencing patterns across domains.

\section{Provocation: Teaching Patterns in the Agent Era}\label{sec:study}

% bring in Pauline's 6 patterns about teaching of the course 

That patterns can be useful in education is well-established.  Here, we  consider whether and how the specific pattern anatomy, along with an adapted version of our discipline for working with patterns, can be enacted and expanded upon by learners and educators.  The setting is Camp CryptoBot, which offers a transformative, mission-driven educational experience designed to introduce high school students to the foundational concepts of cybersecurity.\footnote{\url{https://csis.pace.edu/~mosley/about.html}}

This section can also be read as an exercise in adapting PLACARD, mentioned above, in light of the other ideas described in this paper. For example, rather than teaching CLA’s semantic strata up front (litany/system/worldview/myth), they could be introduced via a loose analogy with the pattern maturity layers, as follows:
\begin{description}
\item[\texttt{:settled} $\leftrightarrow$ litany] These patterns are surface-stable: you can use them without reopening interpretation.
\item[\texttt{:active} $\leftrightarrow$ system] These describe ongoing dynamics, where feedback and coordination costs matter.
\item[\texttt{:greenfield} $\leftrightarrow$ worldview] Competing hypotheses about “how this works” are still in play.
\item[\texttt{:stub} $\leftrightarrow$ myth] Here we are talking about intuitions and speculative frames that have not been operationalized (and may never be).
\end{description}

Our overall thesis is that teaching patterns in the agent era emphasizes practical literacy in recognizing, enacting, and reflecting on patterned action rather than passive content acquisition. This way of working treats each sequence of events in a pattern-aware learning space as an opportunity for further pattern mining.

% Patterns are then introduced through Causal Layered Analysis (CLA) as decision-support tools that harmonize layered choices rather than prescribe linear sequences, with early emphasis on layering as coordination across interactional, organizational, and environmental contexts. CLA remix exercises surface tensions across litany, system, worldview, and myth layers, revealing CLA as a brittle diagnostic frame due to the expertise required for accurate cross-layer mapping. Flexiformal journaling and evidence exercises strengthen context–if–then–because reasoning by requiring enactment, outcome collection, and rationale testing. 

In the context of Camp CryptoBot, quantitative and qualitative feedback revealed recurring pedagogical patterns relevant to cybersecurity education: the largest gains in interest and confidence occurred among students with low prior exposure, indicating that early successful pattern enactment enhances perceived agency. For young women, these scaffolded experiences disrupt self-selection out of cybersecurity by framing competence as situational and learnable. One example of a specific pattern is ``Cooperative Learning as Layered Agency''. We observed that collaborative, team-based learning consistently strengthened confidence and persistence, particularly among female participants. Cooperative environments allow agencies to be layered across individuals and groups, reflecting real-world cybersecurity practice, where decisions are distributed across roles and perspectives. We put forward several patterns in the format outlined in the Introduction, capture key learnings.  To keep the presentation concise, one 
is presented here in its complete form (Figure \ref{fig:cryptobot-example}), and the others are included in
the Appendix.  Taken together, these patterns suggest that agent-era
cybersecurity education is less about early technical mastery and more
about cultivating recognition, enactment, reflection, and identity
within supportive sociotechnical environments.

\begin{figure}
\begin{tcolorbox}[colback=white,colframe=black,boxrule=0.5pt,arc=0.5mm]
\textbullet\  \textbf{summary:}
Experiential activities act less as skill training and more as repeated exposure to decision patterns, allowing learners to recognize applicability conditions without being blocked by abstraction.

\textbullet\ \textbf{context:} Cybersecurity concepts are often abstract, jargon-heavy, and intimidating to new learners, particularly those without prior technical background.

\textbullet\ \textbf{if:} Learners engage in hands-on, experiential activities such as Sphero programming, cipher decoding, or mission-based challenges,

\textbullet\ \textbf{however:} abstract concepts like encryption, confidentiality, and defense in depth remain present and risk overwhelming students if introduced only at a theoretical level,

\textbullet\ \textbf{then:} experiential activities act as pattern recognition drills, training learners to notice when a pattern applies, what conditions trigger it, and what actions follow,

\textbullet\ \textbf{because:} grounding abstract concepts in tangible tasks shifts attention from technical vocabulary to decision-making and enactment,

\textbullet\ \textbf{next-steps:} Examine which cybersecurity patterns are most
reliably recognized through hands-on enactment, and where experiential
grounding begins to break down as technical complexity increases.
\end{tcolorbox}
\caption{Hands-On, Experiential Learning as Pattern Recognition\label{fig:cryptobot-example}}
\end{figure}

\section{Provocation: Working with agents in a Pseudo-Virtual C-Suite}\label{sec:pvcs}
Rather than a company purely run by AI agents—as with TheAgentCompany mentioned in Section \ref{sec:related-work}—one or more humans can stay in the loop. Author Charles Jeffrey Danoff tried this with his company Mr.~Danoff's Teaching Laboratory, LLC. He is the only current employee but he brought in a pseudo virtual C-Suite (PVCS) that included Google Gemini as Chief Operating Officer (COO), Claude as Chief Marketing Officer (CMO), ChatGPT as Chief Ethics and Impact Officer (CEIO), Perplexity as Chief Academic Officer (CAO) and DeepSeek as Chief Revenue Officer (CRO). They co-developed a ninety-day marketing plan to promote a new online bookstore and an extensive renovation of the company homepage. As an example, Mr.~Danoff discussed his goals and available inventory to the CMO. In tandem, they developed a 90 day draft plan which was then shared with the CRO who gave feedback and recommendations to drive revenue including specific updates to the webpage with Google Analytics for tracking clicks to the bookstore. This updated version was sent to the CEIO who reviewed and recommended making the HTML more compliant to accessibility standards. Mr.~Danoff was copy/pasting messages from one LLM to another, manually facilitating the conversation. Following major rounds of activities, agents filled in an adaptation of our PAR (the ``agent Action Review'' or aAR shown in Figure \ref{fig:aar}). The C-Suite has also begun pattern mining for successful ways to work together, such as ``The Sovereign Agency Protocol'' which states ``No agent may speak for another.'' This emerged because one LLM, such as the COO Gemini, frequently made up the responses of others, such as the CRO DeepSeek. By respecting agent sovereignty, consensus across participants is not hallucinated.

\begin{figure}
\begin{tcolorbox}[colback=white,colframe=black,boxrule=0.5pt,arc=0.5mm]
\begin{quote}
\sffamily
Mr. Danoff's Teaching Laboratory, LLC \\
aAR Version 1.1 -- [DATE]

TITLE: [Day \# -- Phase X, Activity Name] \\
DATE / PHASE: [Day \# -- Phase X] (Zero Point: [START DATE]) \\
LEAD AGENTS: [Primary contributor(s)] \\
SUPPORTING AGENTS: [Collaborators/systems]

1. INTENT: What was the collective aim/hypothesis? [Reference plan objective] \\
2. AGENT PERSPECTIVES: What did each agent prioritize/assume? [Constraints/data] \\
3. INTERACTIONS: Where did coordination/friction/synergy occur? \\
4. OUTCOME: Observable results -- quantitative + qualitative. [Metrics/logs] \\
5. LEARNING: Proto-patterns/insights. What worked/failed/adapted? \\
6. NEXT STEPS: Actionable adjustments/data needs/follow-ups. \\
7. SIGN-OFFS: [ ] CEO [ ] COO [ ] CAO [ ] CIO [etc.]A design pattern is a generalized action review, a reusable description of how intentions, actions, and outcomes tend to relate across situations.
\end{quote}
\end{tcolorbox}
\caption{Agent action review (aAR) protocol \label{fig:aar}}
\end{figure}

\section{Conclusion}\label{sec:conclusion}

%% check we don't contradict ourselves
This paper continues our exploration of peeragogy in the presence of AI agents, with a particular focus on practical engagement with contemporary artificial intelligence concepts—most notably the Active Inference Framework. Relative to earlier work in this collaboration, a key step forward in the current project was our ability to condense our methods into a simple form: concise enough to be implemented computationally, while still leaving room for open-ended investigation.  We have shown, at the proof of concept level, that AI systems can be guided by design patterns written in informal language to write software that expresses the authors' intent.

In the context of contemporary automation, we see this work as contributing to a more structured way of thinking about which forms of work, judgment, and creativity remain meaningfully and essentially human. Education is not exempt from the expanding scope of automation and is perhaps uniquely sensitive to its consequences. If AI systems generate assignments, grade submissions, and tutor students, the distinctive value of human participation becomes unclear. Ultimately—as anticipated in Alan Turing’s early speculations \cite{turing1950computing}—the students themselves might be replaced by computer programs, raising big questions not only for universities but for everyone.

Our approach in this paper points toward a non-dystopian synthesis: the possibility that agent-based systems can substantially accelerate and extend human learning, while humans remain actively engaged in the design, evaluation, and governance of computational systems. For this synthesis to be viable, however, the norms of the scientific method must continue to be applied to LLM-mediated work. As Ibn al-Haytham argued, the seeker of truth must question inherited authority and submit claims to argument and demonstration. This requirement remains true—and is arguably even more important—when working with generative systems whose outputs can appear fluent, persuasive, and authoritative while remaining ungrounded.

In our work, both agents and patterns have been instruments within a broader generative system. We see promising directions forward in the practical application of design patterns as a shared, inspectable, and increasingly computational medium.  We presented two key provocations: a pseudo-virtual C-Suite team using a patterned workflow to help a small business decision-maker interrogate strategic assumptions, and a summer camp employing design pattern methods to help students learn cybersecurity first principles.  As educators observe patterns in how students learn, it helps them reach future students more effectively.  There are related broader implications, towards understanding learning patterns for both humans and AI agents, and towards building more effective ways of working together.  In particular, this paper may serve as a prolegomenon to the use of design patterns in AI governance: by making explicit the kind of learning and revision practices that would be required.

\section{Acknowledgements}\label{sec:ack}
We thank our PLoP shepherd Valentino Vrani\'{c} and Writers' Workshop contributors Rebecca Wirfs-Brock, Christian Kohls, An Hikino, and Bess Sadler. Pace University provided funding to support Mary Tedeschi’s participation in the conference. We note the role of LLMs in supporting both ideation and drafting.

\renewcommand\bibname{References}
\renewcommand\refname{References}

\bibliographystyle{ACM-Reference-Format-Journals}
\bibliography{./main.bib}

\clearpage
\appendix

\section*{Appendix}
\section{Further Technical details}

\begin{table}[h]
\begin{tcolorbox}[colback=white,colframe=black,boxrule=0.5pt,arc=0.5mm]
\begin{quote}
{\sffamily\small
Do not use design-pattern structure by default.

Use the design-pattern structure only when the user explicitly asks for a design pattern.

Prior pattern use, topic suitability, or inferred usefulness do not count as either a request or permission.

PRE-PATTERN CHECK

If a pattern might help but was not requested: ask the user whether they want a pattern.

NATURE OF PATTERNS

Treat patterns as latent analytical tools, not presentation templates. Their primary purpose is to name and hold a tension.

PATTERN STRUCTURE:

Required fields (exactly one each, in order): context, if, however, then, because, next-steps.

CONSTRAINTS

the pair if—however outline the tension.

because must give a single primary rationale.

next-steps must not resolve the tension, but should give further evidence that the pattern is valid.}
\end{quote}
\end{tcolorbox}
\caption{System prompt for adding pattern design pattern awareness to, e.g., ChatGPT\label{box:1}}
\end{table}

\begin{table}[b]
\begin{tcolorbox}[colback=white,colframe=black,boxrule=0.5pt,arc=0.5mm]
\begin{center}
\begin{minipage}{.95\textwidth}
{\small
\raggedright
\TabPositions{16em}
Session Start\tabto{16em}// begin cycle\\
AIF Pattern Selector\tabto{16em}// select pattern\\
- Current beliefs ($\mu$)\tabto{16em}// pattern evidence + tau-cache + maturity phase\\
\tabto{1em}\emph{Exploratory trigger:}\tabto{16em}// $\tau < \text{min-sample} \Rightarrow$ explore mode, or explicit flag\\
\tabto{1em}- Exploratory mode add-ons\tabto{16em}// + patterns tried/available + codebase state + uncertainty\\
- Evidence + priors\tabto{16em}// pattern scores\\
- Anchors/forecast\tabto{16em}// intent + plan\\
Sample via softmax($-G/\tau$)\tabto{16em}// for each candidate, compute logit = $-G/\tau$, normalize with softmax \\
\tabto{1em}\emph{Log PSR; switch to explore if $\tau$ low}\tabto{16em}// get a distribution, and sample a pattern from it.\\
Pattern Application\tabto{16em}// via \texttt{pattern-action} RPC (or inferred)\\
\tabto{1em}\emph{Pattern and system constraints:}\tabto{16em}// e.g., selection before implement/update; ``\textbf{+then}...'', ``\textbf{+next-steps}...''
\tabto{1em}- Action generation\tabto{16em}// command\_execution, reasoning, etc.\\
\tabto{1em}- Tool calls\tabto{16em}// observed ops, e.g. \texttt{musn-plan}, \texttt{musn-hud}, etc.\\
Observe outcome\tabto{16em}// deltas: tool outputs, file edits,
  test results,  logged actions\\
Belief Update\tabto{16em}// update evidence counts, recompute $\tau$, and refresh belief summaries\\
\tabto{1em}- Record evidence / error proxy \tabto{16em}//  pattern- \{read, update, implement\} $\Rightarrow$ per-pattern evidence\\
\tabto{1em}- Update policy precision ($\tau$)\tabto{16em}// text length (=error proxy), precision priors, candidate score spread\\
\emph{Log PUR}\tabto{16em}// via observed \texttt{pattern-use} (or inferred)\\
Next tick...\tabto{16em}// continue
}
\end{minipage}
\end{center}
\end{tcolorbox}
\caption{Pseudo-code for \texttt{fulab} agents\label{box:2}}
\end{table}

\begin{table}
\begin{tcolorbox}[colback=white,colframe=black,boxrule=0.5pt,arc=0.5mm]
\begin{tabular}{ll}
\begin{minipage}{.5\textwidth}
{\footnotesize
\begin{verbatim}
@flexiarg p4ng/agent-command-pattern
@title Agent Command Pattern
@audience futon developers, CS students, pattern agents
@tone analytic
@style design-pattern

! conclusion: Encapsulate every such operation as a
 Command value that carries `:id`, parameters,
 provenance, capability tags, and optional `undo!`,
 and run it only through an execution pipeline (local
 worker, remote service, or threaded pool). Commands
 can be persisted, reordered, decorated, and routed
 to specialised executors such as a git agent or math
 agent.

  + context: You are defining a reliable execution
 pipeline for agent actions.

  + if: Agents perform repo scans, graph updates, and
 external API calls by invoking ad hoc functions with
 no shared interface, so actions cannot be queued,
 audited, or retried safely.

  + however: Higher-layer orchestration needs
 side-effectful work to travel through queues, logs,
 schedulers, and undo ledgers without each caller
 reinventing a protocol.

  + then: Encapsulate every such operation as a
 Command value that carries `:id`, parameters,
 provenance, capability tags, and optional `undo!`,
 and run it only through an execution pipeline (local
 worker, remote service, or threaded pool). Commands
 can be persisted, reordered, decorated, and routed
 to specialised executors such as a git agent or math
 agent.

  + because: A uniform command abstraction gives the
 stack a shared action language: histories become
 replayable, undo/redo is tractable, thread pools can
 execute arbitrary work safely, and schedulers can
 move labour across machines without touching
 implementation internals.

  + next-steps: next[Log one concrete instance of
 this pattern in the futon3 ledger.]
\end{verbatim}
}
\end{minipage}
&
\begin{minipage}{.5\textwidth}
{\footnotesize
\begin{verbatim}
@arg fulab/blast-radius

! conclusion: Clearly define the scope of impact
  from failure and provide a minimal rapid
  recovery plan.

  + context: A change or tool action could impact
 multiple subsystems or teams.

  + if: You want failures to remain localized and
 diagnosable without full rollback.

  + however: Without a declared blast radius,
 incident response defaults to broad freezes or
 guesswork.

  + then: Emit a blast-radius event before risky
 actions, naming affected surfaces, rollback scope,
 and detection signals; record the outcome on
 clock-out.

  + because: Declared boundaries shorten diagnosis
 time and reduce collateral damage.

  + next-steps: Require blast-radius events for
 cross-futon changes; audit that the rollback scope
 matches actual artifacts.
\end{verbatim}

\begin{verbatim}
@arg p4ng/timebox-the-core-agent-prime

! instantiated-by: Timebox the Core (Agent-Facing)

  + context: An agent is operating within
 fixed-duration collaborative sessions.

  + if: The agent attempts to complete a full
  inquiry loop.

  + however: Feedback indicates saturation or
 overload before completion.

  + then: Re-focus dynamically on the core loop,
 complete a concise iteration, and earmark unresolved
 deeper insights for a follow-up session.

  + because: Efficiency in the moment and depth
 deferred preserves flow without sacrificing insight.

  + next-steps: Log one concrete instance of this
 pattern in the futon3 ledger.
\end{verbatim}
}
\end{minipage}
\end{tabular}
\end{tcolorbox}
\caption{Examples of design patterns read and written by agents with light markup for readability\label{box:3}}
\end{table}

\FloatBarrier

\section{Additional Camp Cryptobot Patterns}

These patterns expand the discussion from Section \ref{sec:study}.

\hypertarget{mission-based-and-narrative-learning-as-par-studios}{%
\subsubsection*{Mission-Based and Narrative Learning as PAR Studios}\label{mission-based-and-narrative-learning-as-par-studios}}
~\newline\hspace{.8em} \textbullet\ \textbf{summary:} Narrative missions reframe cybersecurity tasks as exploratory cycles, creating conditions where uncertainty and failure contribute to learning rather than undermining confidence.

\textbullet\ \textbf{context} Fear of failure and evaluation pressure can inhibit
experimentation and learning in cybersecurity education.

\textbullet\ \textbf{if} Challenges are framed as mission-driven, narrative scenarios
such as fictional cybersecurity breaches,

\textbullet\ \textbf{however} students still confront uncertainty, adversarial
reasoning, and incomplete information,

\textbullet\ \textbf{then} missions operate as PAR studios that normalize iteration
and support reflection on intention, outcome, and surprise.

\textbullet\ \textbf{because} narrative framing shifts tasks from high-stakes
evaluation to exploration of a problem space where failure is treated as
informative.

\textbullet\ \textbf{next-steps} Investigate how much narrative scaffolding is
required to sustain experimentation without obscuring the underlying
technical and strategic patterns.

\hypertarget{cooperative-learning-as-layered-agency}{%
\subsubsection*{Cooperative Learning as Layered
Agency}\label{cooperative-learning-as-layered-agency}}

~\newline\hspace{.8em} \textbullet\ \textbf{summary:} Team-based learning environments distribute agency across participants, aligning educational practice with the inherently collective nature of real-world cybersecurity work.

\textbullet\ \textbf{context} Cybersecurity practice in the real world is
distributed, collaborative, and role-differentiated rather than
individually linear.

\textbullet\ \textbf{if} learning environments emphasize cooperative, team-based
problem solving,

\textbullet\ \textbf{however} individual learners may still experience isolation,
confidence gaps, or stereotype threat,

\textbullet\ \textbf{then} agency becomes layered across individuals and groups,
strengthening confidence, persistence, and shared ownership of outcomes.

\textbullet\ \textbf{because} distributed decision-making mirrors real-world
cybersecurity practice and provides peer support that reduces social and
cognitive load.

\textbullet\ \textbf{next-steps} Clarify which forms of role differentiation and
coordination best support layered agency without reintroducing hierarchy
or exclusion.

\hypertarget{role-models-and-career-pathways-as-worldview-reframing}{%
\subsubsection*{Role Models and Career Pathways as Worldview
Reframing}\label{role-models-and-career-pathways-as-worldview-reframing}}

~\newline\hspace{.8em} \textbullet\ \textbf{summary:} Visible role models intervene at the level of narrative and identity, reshaping learners’ assumptions about who can participate in cybersecurity and why.

\textbullet\ \textbf{context} Perceptions of who ``belongs'' in cybersecurity shape
participation, motivation, and long-term engagement.

\textbullet\ \textbf{if} students are exposed to guest speakers and female
cybersecurity professionals sharing firsthand accounts,

\textbullet\ \textbf{however} existing cultural narratives about identity,
capability, and legitimacy remain deeply embedded,

\textbullet\ \textbf{then} these encounters reframe learners' worldviews by expanding
their sense of possible futures in the field.

\textbullet\ \textbf{because} pattern operation extends beyond interactional skills
into worldview and myth layers that shape aspiration and
self-identification.

\textbullet\ \textbf{next-steps} Explore how often and in what forms such
worldview-level interventions must occur to counteract dominant cultural
narratives.

\hypertarget{low-barrier-entry-as-early-pattern-enactment}{%
\subsubsection*{Low-Barrier Entry as Early Pattern
Enactment}\label{low-barrier-entry-as-early-pattern-enactment}}

~\newline\hspace{.8em} \textbullet\ \textbf{summary:} Low-barrier tools enable early enactment of core cybersecurity logic, allowing confidence and pattern understanding to develop before technical complexity escalates.

\textbullet\ \textbf{context} Many learners enter cybersecurity education with little
or no prior coding experience.

\textbullet\ \textbf{if} instruction begins with low-barrier tools and scaffolded
platforms such as Sphero,

\textbullet\ \textbf{however} learners will eventually need to confront more complex
technical environments,

\textbullet\ \textbf{then} early success enables pattern enactment focused on logic,
sequencing, and security principles rather than syntax.

\textbullet\ \textbf{because} validating the if--then--because structure of
cybersecurity patterns builds self-efficacy that supports later
complexity.

\textbullet\ \textbf{next-steps} Determine where and how transitions to
higher-barrier tools can occur without undermining early confidence.

\hypertarget{reflection-as-pattern-articulation-and-identity-formation}{%
\subsubsection*{Reflection as Pattern Articulation and Identity
Formation}\label{reflection-as-pattern-articulation-and-identity-formation}}

~\newline\hspace{.8em} \textbullet\ \textbf{summary:} Structured reflection stabilizes learning by making patterns explicit and by linking technical practice to ethical reasoning and emerging professional identity.

\textbullet\ \textbf{context} Learning gains remain fragile if experiences are not
consolidated and integrated into a learner's sense of self.

\textbullet\ \textbf{if} students engage in structured reflection through end-of-day
reflections and mission debriefs,

\textbullet\ \textbf{however} reflection requires time, guidance, and institutional
commitment,

\textbullet\ \textbf{then} reflection functions as pattern articulation, supporting
metacognition, ethical reasoning, and emerging professional identity.

\textbullet\ \textbf{because} explicit articulation of context, action, outcome, and
rationale stabilizes learning and connects practice to
self-understanding.

\textbullet\ \textbf{next-steps} Assess which reflective prompts most effectively
surface pattern understanding without turning reflection into rote
reporting.

\end{document}